\documentclass{elsart}
 \usepackage{graphics}
\begin{document}
\begin{frontmatter}

\title{Chiral condensates and size of the sigma term}
\author{G.\ X.\ Peng}
  \ead{gxpeng@lns.mit.edu, gxpeng@ihep.ac.cn}
\address[MIT]{%
 Center for Theoretical Physics,
 Lab.\ for Nuclear Science and Depart.\ of Physics, \\
 Massachusetts Institute of Technology,
 77 Massachusetts Avenue, Cambridge,\\
 MA 02139-4307, USA
             }
\address[IHEP]{Institute of High Energy Physics,
               Chinese Academy of Sciences,
               19B Yuquanlu, Beijing 100039, China}

\begin{abstract}
The in-medium chiral condensate is studied with a new approach
which has the advantage of no need for extra assumptions on the
current mass derivatives of model parameters.
It is shown that the pion-nucleon sigma term is 9/2 times the
average current mass of light quarks, if quark confinement is linear.
Considering both perturbative and non-perturbative interactions,
the chiral condensate decreases monotonously with increasing densities,
approaching to zero at about 4 fm$^{-3}$.
\end{abstract}

\end{frontmatter}

\newpage

The calculation of in-medium quark condensates is of
crucial importance to the chiral property of
quantum chronmodynamics (QCD)
%\cite{Brown02PR363,Colangelo01PRL86}.
\cite{Brown02PR363}.
A popular method for this is the Feynman-Helmann theorem
which gives \cite{Cohen92PRC45}
\begin{equation} \label{qcHF}
\frac{\langle\bar{q}q\rangle_{n_{\mathrm{b}}}}
     {\langle\bar{q}q\rangle_0}
=1-\frac{1}{2|\bar{q}q|_0}
   \frac{\partial E}{\partial m_0},
\end{equation}
where $E$ is the energy density above the vacuum,
$m_0$ is the current mass of $u/d$ quarks,
and
$
|\bar{q}q|_0
\equiv |\langle\bar{q}q\rangle_0|
=-\langle\bar{q}q\rangle_0
\approx (225\ \mbox{MeV})^3.
$
Eq.\ (\ref{qcHF}) can be schematically derived as such.
From Feynman-Helmann theorem one has
$
\langle\Psi(\lambda)|\frac{d}{d\lambda}H(\lambda)|\Psi(\lambda)\rangle
=\frac{d}{d\lambda}\langle \Psi(\lambda)|H(\lambda)|\Psi(\lambda)\rangle
$
for the hamiltonian $H(\lambda)$ with the eigenstate
$|\Psi(\lambda)\rangle$. Write the QCD
hamiltonian density for the two flavor symmetric case
as $H_{\mathrm{QCD}}=H^{\prime}+2m_0\bar{q}q$
where the second term breaks the chiral symmetry explicitly while the
first term is the remaining part which has nothing to do with the
current mass $m_0$. Then making the substitution $\lambda\rightarrow m_0$
and $H\rightarrow \int d^3xH_{\mathrm{QCD}}$ gives
$
2\langle\Psi(m_0)|\bar{q}q|\Psi(m_0)\rangle
=\frac{\partial}{\partial m_0}
\langle\Psi(m_0)|H_{\mathrm{QCD}}|\Psi(m_0)\rangle.
$
Here the integration over space is canceled due to
the uniformity of the system. The derivative has been changed to
partial derivative because the system energy may depend on other
independent quantities for example the density $n_{\mathrm{b}}$.
Appply the equality just obtained to the state
$
|\Psi\rangle=|n_{\mathrm{b}}\rangle
$
and
$|\Psi\rangle=|0\rangle$,
and then taking the difference lead to Eq.\ (\ref{qcHF}) naturally.

Because no one can exactly solve QCD presently, the energy
density $E$ in Eq.\ (\ref{qcHF}) is given with some model
parameters. The main difficulty in this formula is that one
has to know the current mass derivatives of model parameters
which are, except for a few special cases,
usually not available.

Recently, another approach has been proposed
 \cite{Peng02PLB548,Peng03IJMPA18} with
%It bypasses the difficulty via introducing an equivalent mass, and
%therefore has
the advantage of no need for assumptions on
the current mass derivatives of model parameters.
However, how the
effective Fermi momentum is connected to density was not
investigated clearly. In Ref.~\cite{Peng02PLB548}, it is merely
boosted by a simple factor of 2
while not boosted in the section 2 of Ref.\ \cite{Peng03IJMPA18},
compared with the non-interacting case.
In this paper, it will be generally proved that the
relation between the effective Fermi momentum and density is
determined by general principles of thermodynamics.
At lower densities, a simple relation among the pion-nucleon
sigma term, the quark current mass, and quark confinement
is found, which means that, if the confinement is linear,
the pion-nucleon sigma term should be 9/2 times the average
current mass of light quarks.
Full density behavior of the in-medium chiral condensate
is calculated by considering both perturbative and
non-perturbative interaction effects,
which gives a zero condensate at about 4 fm$^{-3}$.

Let's outline the key points of the new approach in
Refs.~\cite{Peng02PLB548,Peng03IJMPA18} for the two flavor case.
Ignoring terms breaking flavor symmetry, the corresponding
QCD hamiltonian density can be schematically written as
$% \begin{equation}     \label{Hqcd}
  H_{\mathrm{QCD}}
  =H_{\mathrm{k}}
   + 2 m_{0}\bar{q}q
   + H_{\mathrm{I}},
$  % \end{equation}
where $H_{\mathrm{k}}$ is the kinetic term, $H_{\mathrm{I}}$ is the
interacting part,
and $m_0$ is the average current mass of $u$ and $d$ quarks.
Define a new operator
$%\begin{equation}    \label{Hequiv}
  H_{\mathrm{eqv}}\equiv H_{\mathrm{k}} + 2 m \bar{q}q
$ %\end{equation}
with $m$ being an equivalent mass
% (referred to as an equivalent mass to avoid confusing with
% the normal effective mass)
to be determined by the requirement
that
%
% $H_{\mathrm{eqv}}$ has the same expectation
%value with $H_{\mathrm{QCD}}$ for any state $|\Psi\rangle$, i.e.
%$
%  \langle{\Psi} | H_{\mathrm{eqv}} |\Psi\rangle
% =\langle{\Psi} | H_{\mathrm{QCD}} |\Psi\rangle.
%$\
%Applying this equality to the state $|n_{\mathrm{b}}\rangle$\
%% with baryon number density $n_{\mathrm{b}}$
%and to the vacuum $|0\rangle$, respectively,
%taking then the difference, one has
%
it meets
$ %\begin{equation} \label{Hrel}
  \langle H_{\mathrm{eqv}}\rangle_{n_{\mathrm{b}}}
 -\langle H_{\mathrm{eqv}}\rangle_0
 =\langle H_{\mathrm{QCD}}\rangle_{n_{\mathrm{b}}}
 -\langle H_{\mathrm{QCD}}\rangle_0.
$ %\end{equation}

   Since particles considered are uniformly distributed,
or in other words, $n_{\mathrm{b}}$ has
nothing to do with space coordinates, one can write
$
\langle \Psi|m(n_{\mathrm{b}})\bar{q}q|\Psi\rangle
=m(n_{\mathrm{b}})\langle \Psi|\bar{q}q|\Psi\rangle.
$
This equality is especially obvious if it is considered in terms
of quantum mechanics: $|\Psi\rangle$ is a wave function with
arguments $n_{\mathrm{b}}$ and coordinates, the expectation value is
nothing but an integration with respect to the coordinates.
Therefore, if $n_{\mathrm{b}}$ does not depend on coordinates, the function
$m(n_{\mathrm{b}})$ is also a coordinate-independent c-number, and can
naturally be taken out of the integration.
However, if $n_{\mathrm{b}}$ is local, the case becomes
much more complicated and it will not be considered here.
Now substituting the expressions of $H_{\mathrm{QCD}}$
and $H_{\mathrm{eqv}}$
%Eqs.\ (\ref{Hqcd}) and (\ref{Hequiv})
into the equation
$ %\begin{equation} \label{Hrel}
  \langle H_{\mathrm{eqv}}\rangle_{n_{\mathrm{b}}}
 -\langle H_{\mathrm{eqv}}\rangle_0
 =\langle H_{\mathrm{QCD}}\rangle_{n_{\mathrm{b}}}
 -\langle H_{\mathrm{QCD}}\rangle_0
$ %\end{equation}
leads to
\begin{equation} \label{mdef}
m = m_0+\frac{\langle H_{\mathrm{I}}\rangle_{n_{\mathrm{b}}}
                   -\langle H_{\mathrm{I}}\rangle_0}
{2(\langle\bar{q}q\rangle_{n_{\mathrm{b}}}-\langle\bar{q}q\rangle_0)}
  \equiv m_0 + m_{\mathrm{I}}.
\end{equation}

Therefore, considering quarks as a free system, i.e.,
without interactions,  while keeping the
system energy unchanged, quarks should acquire an
equivalent mass of the form shown in Eq.\ (\ref{mdef}).
From this equation one can see that the
equivalent mass $m$ includes two parts: one is the original mass
or current mass $m_{0}$, the other one is the interacting part
$m_{\mathrm{I}}$.
Obviously the equivalent mass is a function of
both the quark current mass and the density. At finite temperature,
it depends also on the temperature as well. % \cite{Surk}.
%Here only zero temperature is considered.
%
%Due to the quark confinement and asymptotic freedom,
%one may naturally expect
%\begin{equation} \label{mlim}
%\lim_{n_{\mathrm{b}}\rightarrow 0} m_{\mathrm{I}}=\infty\
%  \ \mbox{and}\ \
%\lim_{n_{\mathrm{b}}\rightarrow\infty} m_{\mathrm{I}} =0.
%\end{equation}
%
Because the hamiltonian density $H_{\mathrm{eqv}}$ has the same form
as that of a system of free particles with a density dependent mass $m$,
the corresponding dispersion relation $\varepsilon=\sqrt{p^2+m^2}$
is also density dependent.
The energy density of quark system can then be expressed as
\begin{equation}   \label{epsilon}
E
= \frac{g}{2\pi^2} \int^{p_{\mathrm{f}}}_0
   \sqrt{p^2+m^2}\ p^2 dp
=\frac{gp_{\mathrm{f}}^3}{6\pi^2}
 m F\left(\frac{p_{\mathrm{f}}}{m}\right),
\end{equation}
where % $g$ = 2(flavor) $\times$ 3(color) $\times$ 2(spin) = 12
         $g=12$
is the degeneracy factor, and the function $F$ is defined to be
$
F(x) \equiv \frac{3}{8} [
    x\sqrt{x^2+1} (2x^2+1)-\mbox{sh}^{-1}(x)
                        ]/x^3
$
with
$
\mbox{sh}^{-1}(x)=\mbox{ln}(x+\sqrt{x^2+1}).
$
The effective Fermi momentum $p_{\mathrm{f}}$ is related to
the chemical potential $\mu$\ by
\begin{equation} \label{pfmu}
p_{\mathrm{f}}=\sqrt{\mu^2-m^2}
\ \mbox{or} \
\mu=\sqrt{p_{\mathrm{f}}^2+m^2}.
\end{equation}

The physical meaning of the equivalent mass is that quarks should
have the equivalent mass if the system is free but with unchanged energy.
It is not difficult to prove that such an equivalent mass always exists.
In principle, if one obtains the energy density $E$ from some realistic models
or even from QCD in the future, the equivalent mass $m$ can be obtained by
solving Eq.\ (\ref{epsilon}).
From the point of view of thermodynamics, one can alway
choose to express some quantity as an ideal gas form with
density and/or temperature dependent particle mass(es)
while all other quantities determined by thermodynamic principles.
For example, in Refs.\ \cite{Goloviznin1993ZPC57,Peshier1994PLB337},
the pressure of gluon plasma
is expressed as the same form with that of an ideal gas while the gluon
`thermo mass' $m(T)$ is determined by fitting to lattice data.
In Ref.\ \cite{Gorenstein1995PRD52}, the entropy is expressed
as the same form as that of an ideal gas. Here the energy has been
expressed as the same form as that of an ideal quark gas.
However, how the Fermi momentum $p_{\mathrm{f}}$ is connected to density should
be studied carefully. If no interactions are accounted
or a specific duality principle is assumed, the Fermi momentum
will be connected to density simply by
$p_{\mathrm{f}}=[(18/g)\pi^2n_{\mathrm{b}}]^{1/3}$.
Here quark interactions are included within the equivalent mass.
So its relation to density should be determined by general
thermodynamic principles. It can be shown that
the baryon number density $n_{\mathrm{b}}$ is related to
the Fermi momentum $p_{\mathrm{f}}$ by
\begin{equation}  \label{nbexp}
n_{\mathrm{b}}
=\frac{gp_{\mathrm{f}}^3}{18\pi^2}
 +\frac{g}{12\pi^2}
  \int \left[
 p_{\mathrm{f}}
 -\frac{m^2\mbox{sh}^{-1}(p_{\mathrm{f}}/m)}
       {\sqrt{p_{\mathrm{f}}^2+m^2}}
       \right] mdm.
\end{equation}
%with
%$\mbox{sh}^{-1}(x)=\ln(x+\sqrt{x^2+1})$
%being the hyperbolic sine function.

  To prove this, let's write the fundamental
thermodynamic relation
$ %\begin{equation}  \label{therm1}
d(VE)=Td(VS)-PdV+3\mu d(Vn_{\mathrm{b}})
$ %\end{equation}
which is the combination of the first and second laws of
thermodynamics. Here $P$ is the pressure,
$S$ is the entropy density, and $V$ is the volume.
Now rewrite the equality % Eq.\ (\ref{therm1})
 as
$ %\begin{equation}  \label{therm2}
dE=TdS-(P+E-TS-3\mu n_{\mathrm{b}}){dV}/{V}+3\mu dn_{\mathrm{b}}
$ %\end{equation}
which implies
%\begin{eqnarray}
%%&& T=\left.\frac{dE}{dS}\right|_{n_{\mathrm{b}}},  \label{Tlab}  \\
%&& P+E-TS-3\mu n_{\mathrm{b}}
% =-V\left.\frac{dE}{dV}\right|_{S,n_{\mathrm{b}}}=0,
%                                \label{Plab} \\
%&& \mu=\frac{1}{3}
%       \left.\frac{dE}{dn_{\mathrm{b}}}\right|_S. \label{mulab}
%\end{eqnarray}
$ %\begin{equation}
P+E-TS-3\mu n_{\mathrm{b}}=0\
% \mu=\frac{1}{3} [{dE}/{dn_{\mathrm{b}}}]_S.
\mbox{and}\
\partial E/\partial n_{\mathrm{b}}=3\mu.
$ %\end{equation}
At zero temperature, the entropy is zero (the third law of
thermodynamics),
%Eqs.\ (\ref{Plab}) and (\ref{mulab})
these two equalities become
%\begin{eqnarray}
%P&=&-E+3\mu n_{\mathrm{b}}, \label{Pexp} \\
%dn_{\mathrm{b}}
%&=&{dE}/{(3\mu)}.   \label{dn1}
%\end{eqnarray}
\begin{equation} \label{Pnb}
P=-E+3\mu n_{\mathrm{b}}, \ \
dn_{\mathrm{b}}
={dE}/{(3\mu)}.
\end{equation}

In the present approach, the energy density
is given in Eq.\ (\ref{epsilon}) from which one has
$ %\begin{equation} \label{dE}
dE=\frac{\partial E}{\partial p_{\mathrm{f}}} dp_{\mathrm{f}}
   +\frac{\partial E}{\partial m}dm.
$ %\end{equation}
%with
%$
%{\partial E}/{\partial m}
% = \frac{gm}{4\pi^2}
%             [
% p_{\mathrm{f}}\sqrt{p_{\mathrm{f}}^2+m^2}
%-m^2\mbox{arcsinh} ({p_{\mathrm{f}}}/{m})
%             ]
%$
%and
%$
%{\partial E}/{\partial p_{\mathrm{f}}}
%=\frac{g}{2\pi^2}
%   p_{\mathrm{f}}^2\sqrt{p_{\mathrm{f}}^2+m^2}.
%$
Substituting
% Eq.\ (\ref{dE})
this
 into the second equality of
Eq.\ (\ref{Pnb}) and then
%\begin{equation}  \label{dn2}
%dn_{\mathrm{b}}
%=\frac{gp_{\mathrm{f}}^2}{6\pi^2} dp_{\mathrm{f}}
%  +\frac{gm}{12\pi^2}
%     \left[
% p_{\mathrm{f}}-\frac{m^2\mbox{arcsinh}(p_{\mathrm{f}}/m)}
%                     {\sqrt{p_{\mathrm{f}}^2+m^2}}
%    \right] dm
%\end{equation}
integrating over both sides will give Eq.\ (\ref{nbexp}) naturally.

Defining
$
 E_{\mathrm{I}}
\equiv
\langle H_{\mathrm{I}}\rangle_{n_{\mathrm{b}}}
-\langle H_{\mathrm{I}}\rangle_0,
$
%and
%$
% n^*\equiv -(2/3)\langle\bar{q}q\rangle_0
%  = {M_\pi^2F_\pi^2}/{(3m_0)} \approx 0.98\ \mbox{fm}^{-3}
%$
%with $M_{\pi}\approx 140$ MeV being the pion mass and
%$F_{\pi}=93.2$ being the pion decay constant,
the interacting part of the equivalent mass,
 $m_{\mathrm{I}}$ in Eq.\ (\ref{mdef}), can be re-written as
$ %\begin{equation} \label{mI2}
m_{\mathrm{I}}={E_{\mathrm{I}}/(3n^*)}
              /({1-\langle\bar{q}q\rangle_{n_{\mathrm{b}}}
                /\langle\bar{q}q\rangle_0})
$ %\end{equation}
with $n^*\equiv (2/3)|\bar{q}q|_0$.
Solving for the ratio
$\langle\bar{q}q\rangle_{n_{\mathrm{b}}}/\langle\bar{q}q\rangle_0$,
this equation leads to
\begin{equation} \label{qc15}
\frac{\langle\bar{q}q\rangle_{n_{\mathrm{b}}}}
     {\langle\bar{q}q\rangle_0}
=1-\frac{E_{\mathrm{I}}}{3n^*m_{\mathrm{I}}}.
\end{equation}

According to % Eq.\ (\ref{Hqcd}),
the expression of $H_{\mathrm{QCD}}$,
the total energy density of
the quark system can be expressed as
\begin{equation} \label{EEI}
E
=\frac{g}{2\pi^2}\int_0^{p_{\mathrm{f}0}}
     \sqrt{p^2+m_0}p^2 dp
 +E_{\mathrm{I}}
\equiv E_0+E_{\mathrm{I}}.
\end{equation}
The first term is the energy density without interactions,
the second term $E_{\mathrm{I}}$ is the contribution from interactions,
and the Fermi momentum $p_{\mathrm{f}0}$ for the non-interacting
case is connected to density by
$ %\begin{equation} \label{pf0nb}
p_{\mathrm{f}0}
=\left[
(18\pi^2/g)
 n_{\mathrm{b}}
 \right]^{1/3}.
$ %\end{equation}

On the other hand,
$E$ has already been expressed in Eq.\ (\ref{epsilon}).
So, replacing the $E$ on the left hand side of Eq.~(\ref{EEI}) with the
right hand side of
Eq.\ (\ref{epsilon}), then dividing by $3n_{\mathrm{b}}$,
one has
\begin{equation} \label{EIexp}
\left(\frac{p_{\mathrm{f}}}{p_{\mathrm{f}0}}\right)^3
  mF\left(\frac{p_{\mathrm{f}}}{m}\right)
 -m_0F\left(\frac{p_{\mathrm{f}0}}{m_0}\right)
=\frac{E_{\mathrm{I}}}
      {3n_{\mathrm{b}}}.
\end{equation}

Substituting Eq.\ (\ref{EIexp}) into Eq.\ (\ref{qc15}) leads to
\begin{equation} \label{qc2}
\frac{\langle\bar{q}q\rangle_{n_{\mathrm{b}}}}
     {\langle\bar{q}q\rangle_0}
=1-\frac{n_{\mathrm{b}}}{n^*m_{\mathrm{I}}}
   \left[
%   \left(
   \frac{p_{\mathrm{f}}^3}{p_{\mathrm{f}0}^3}
%   \right)^3
   mF\left(\frac{p_{\mathrm{f}}}{m}\right)
  -m_0F\left(\frac{p_{\mathrm{f}0}}{m_0}\right)
   \right].
\end{equation}

At the same time,
%combining Eqs.\ (\ref{Pnb}) and (\ref{EEI}) or
taking the derivative with respect to $n_{\mathrm{b}}$
on both sides of
 Eq.\ (\ref{epsilon})
% Eq.\ (\ref{EIexp})
gives
\begin{equation}  \label{dEIexp}
\sqrt{p_{\mathrm{f}}^2+m^2}
%% -\sqrt{p_{\mathrm{f}0}^2+m_0^2}
=\frac{1}{3}
 \frac{dE}{dn_{\mathrm{b}}}.
\end{equation}

Therefore, if one knows the energy density
$E$,
%% $E_{\mathrm{I}}$,
the equivalent mass $m$ and effective Fermi
momentum $p_{\mathrm{f}}$ can be obtained by solving the
Eqs.\ (\ref{dEIexp}) and (\ref{epsilon}),
and the chiral condensate can then be
calculated from Eq.~(\ref{qc2}). The great advantage of this
scheme to calculate the in-medium chiral condensate is that
one does not need to make any assumption on the
current mass derivatives of model parameters.
In fact, the current mass dependence can be derived like this.
Substituting Eq.\ (\ref{epsilon}) into Eq.\ (\ref{qcHF}),
performing the derivative with respect to $m$ at fixed $p_{\mathrm{f}}$,
then comparing with Eq.\ (\ref{qc2}), one will have
\begin{equation}
\frac{\partial m}{\partial m_0}
=\frac{mF(p_{\mathrm{f}}/m)
       -(p_{\mathrm{f0}}/p_{\mathrm{f}})^3
       m_0F(p_{\mathrm{f0}}/m_0)
       }
      {m_{\mathrm{I}}f(p_{\mathrm{f}}/m)}
\end{equation}
with
$
f(x) \equiv
\frac{3}{2}[x\sqrt{x^2+1}-\ln(x+\sqrt{x^2+1})]/x^3.
$

%
%Equation\ (\ref{mdef})
%does not mean the full complexity of the non-abelian QCD
%interaction can be integrated out and encoded only in the
%equivalent mass.
%It is obtained from a definition which ensures
%that the total energy density be written in the same form as
%that of a non-interacting system with quarks carrying the
%equivalent mass, and such a mass uniquely exists.
%However, it is still a happy
%thing to check the validity of the present formalism,
%although no extra assumptions has been introduced.
%

%Now let's try to check the validity of the present formalism.
%
%First, check the extreme case of $m_{\mathrm{I}}=0$
%which is for a non-interacting quark system. In this case,
%Eq.~(\ref{nbexp}) gives
%$
%p_{\mathrm{f}}
%=(18\pi^2n_{\mathrm{b}}/g)^{1/3}
%=p_{\mathrm{f}0}.
%$
%Accordingly, taking the limit of $m_{\mathrm{I}}\rightarrow 0$
%on the right hand side of Eq.~(\ref{qc2}) gives
%$
%{\langle\bar{q}q\rangle_{n_{\mathrm{b}}}}
%     /{\langle\bar{q}q\rangle_0}
%=1-({n_{\mathrm{b}}}/{n^*})
%   f\left({p_{\mathrm{f}0}}/{m_0}\right).
$ $
%  % Eq.~(\ref{qcfree})
%This is an exactly correct result because it
%can be obtained directly from the Feynman-Helmann theorem
%by substituting the $E_0$ in Eq.\ (\ref{EEI}) into Eq.\ (\ref{qcHF}).

Let's compare the present equation of state
to the famous MIT bag model.
According to Eqs.\ (\ref{Pnb}), (\ref{epsilon}),
and (\ref{nbexp}), the pressure of a cold quark plasma is
\begin{eqnarray} \label{Pmu}
P&=& \frac{gm^4}{48\pi^2}\left[
    \frac{\mu}{m}\sqrt{\frac{\mu^2}{m^2}-1}\left(2\frac{\mu^2}{m^2}-5\right)
    +3\mbox{ch}^{-1}\left(\frac{\mu}{m}\right)
                    \right]  \nonumber\\
 & &
    +\frac{g\mu}{4\pi^2} \int
    \left[
     \sqrt{\mu^2-m^2}
     -\frac{m^2}{\mu}\mbox{ch}^{-1}\left(\frac{\mu}{m}\right)
    \right] m dm,
\end{eqnarray}
where $\mbox{ch}^{-1}(x)\equiv\ln(x+\sqrt{x^2-1})$.
For comparison purpose, the independent state variable
has been changed to the chemical potential $\mu$
through Eq.~(\ref{pfmu}).
Obviously, the second term is
the correspondence of the bag constant $B$. The difference is
that the bag constant here is not really a constant. Instead,
it is density-dependent.

Because baryon masses depend on the in-medium chiral condensate,
the total energy of the system also depends on it \cite{Ioffe04}.
In principle, the hadronic matter bellow chiral phase transition
point, where the quark condensate is nonzero, should be described
in terms of hadronic degree of freedom. Otherwise,
both perturbative and nonperturbative quark interactions, including
quark confinement, should be accounted \cite{Ioffe93PRD47}.
So let's expand the equivalent mass to Laurent series
($m_{\mathrm{I}}$ must be divergent at $p_{\mathrm{f}}=0$
or $n_{\mathrm{b}}=0$ due to quark confinement), and merely take
the leading terms in both directions:
\begin{equation} \label{mIexpn}
m_{\mathrm{I}}
=\frac{a_{-1}}{p_{\mathrm{f}}}
% +m_0
 +a_{1} p_{\mathrm{f}}.
\end{equation}
In the following, it will be seen that
the first term is non-perturbative, mainly originated from
the linear confinement of quarks, and the second term is
from the leading contribution of perturbative interactions
with the coefficient $a_{1}$ connected
to the QCD coupling $\alpha_{\mathrm{s}}$ by
\begin{equation} \label{C1alphas}
a_{1}=\sqrt{{2\alpha_{\mathrm{s}}}
                 /({\pi-2\alpha_{\mathrm{s}}})}.
\end{equation}

It is not possible to compare the present formalism
at finite density with lattice results.
However, there are several expressions
for the pressure of a cold quark plasma,
e.g., those from the hard-thermal-loop
perturbation theory \cite{Baier00PRL} and from the
weak-coupling expansion
\cite{Freedman78PRD,Toimela85IJTP}.
Although they are different in higher orders,
their leading term is identical.
%
%With the modified minimal subtraction renormalization
%scheme, the perturbative quark thermal mass
%is given in leading order
%of $\alpha_{\mathrm{s}}$ by
%$m_{\mathrm{f}}=\sqrt{2\alpha_{\mathrm{s}}/(3\pi)}\mu$.
%Motivated by this,
Let's assume the interacting equivalent
mass $m_{\mathrm{I}}$ is, at the perturbative densities and
to leading order, also proportional to the chemical
potential $\mu$, i.e.,
$%\begin{equation}  \label{mIpert}
m_{\mathrm{I}}
=\alpha_0 \mu.
$\ %\end{equation}
It is shown in the following that the coefficient
is $\alpha_0=\sqrt{2\alpha_{\mathrm{s}}/\pi}$.
In fact, the pressure in Eq.~(\ref{Pmu})
can be expanded to Taylor series at $m_{\mathrm{I}}=0$,
i.e.,
%\begin{widetext}
%\begin{eqnarray} \label{Pexpand}
%P&=&P_{\mathrm{id}}
% -\frac{gm_0}{4\pi^2} \left\{
%                      \left[
%\mu\sqrt{\mu^2-m_0^2}-m_0^2\mbox{arccosh}\left(\frac{\mu}{m_0}\right)
%                     \right] m_{\mathrm{I}}
% + %  \frac{gm_0}{4\pi^2}
%         \mu \int\left[
%\sqrt{\mu^2-m_0^2}
%+\frac{m_0^2}{\mu}\mbox{arccosh}\left(\frac{\mu}{m_0}\right)
%                      \right] dm_{\mathrm{I}}
%                      \right\}      \nonumber\\
%&& \hspace{-0.5cm}
%   - \frac{g}{8\pi^2} \left\{
%                      \left[
%\mu\sqrt{\mu^2-m_0^2}-3m_0^2\mbox{arccosh}\left(\frac{\mu}{m_0}\right)
%                   \right] m_{\mathrm{I}}^2
% +  %   \frac{g}{4\pi^2}
%        \mu \int\left[
%\frac{\mu^2-3m_0^2}{\sqrt{\mu^2-m_0^2}}
%+3\frac{m_0^2}{\mu}\mbox{arccosh}\left(\frac{\mu}{m_0}\right)
%                     \right]
%       dm_{\mathrm{I}}^2
%                     \right\}
%+\dots
%\end{eqnarray}
%\end{widetext}
%
\begin{eqnarray} \label{Pexpand}
P&=&
 P_{\mathrm{id}}
 -\frac{gm_0}{4\pi^2} \left\{
                      \left[
\mu\sqrt{\mu^2-m_0^2}-m_0^2\mbox{ch}^{-1}\left(\frac{\mu}{m_0}\right)
                     \right] m_{\mathrm{I}}
                      \right.\nonumber\\
&&
  \left. - %  \frac{gm_0}{4\pi^2}
         \mu \int\left[
\sqrt{\mu^2-m_0^2}
-\frac{m_0^2}{\mu}\mbox{ch}^{-1}\left(\frac{\mu}{m_0}\right)
                      \right] dm_{\mathrm{I}}
                      \right\}      \nonumber\\
&&
   - \frac{g}{8\pi^2} \left\{
                      \left[
\mu\sqrt{\mu^2-m_0^2}-3m_0^2\mbox{ch}^{-1}\left(\frac{\mu}{m_0}\right)
                   \right] m_{\mathrm{I}}^2
                  \right. \nonumber\\
&&
  \left.
-  %   \frac{g}{4\pi^2}
        \mu \int\left[
 % \frac{\mu^2-3m_0^2}{\sqrt{\mu^2-m_0^2}}
  \sqrt{\mu^2-m_0^2}
-3\frac{m_0^2}{\mu}\mbox{ch}^{-1}\left(\frac{\mu}{m_0}\right)
                \right]
       dm_{\mathrm{I}}^2
                     \right\}
                 \nonumber\\
&& +\ \mbox{higher order terms}\ %  m_{\mathrm{I}}
\end{eqnarray}
%
%\begin{eqnarray} \label{Pexpand}
%P&=&P_{\mathrm{id}}
% -\frac{gm_0^3}{4\pi^2} \left\{
%                      \left[
%y\sqrt{y^2-1}-\mbox{ch}^{-1}(y)
%                     \right] m_{\mathrm{I}}
%                      \right.\nonumber\\
%&&\left.  + %  \frac{gm_0}{4\pi^2}
%         y \int\left[
%\sqrt{y^2-1}
%+\mbox{ch}^{-1}(y)/y
%                      \right] dm_{\mathrm{I}}
%                      \right\}      \nonumber\\
%&& %\hspace{-0.5cm}
%   - \frac{gm_0^2}{8\pi^2} \left\{
%                      \left[
%y\sqrt{y^2-1}-3\mbox{ch}^{-1}(y)
%                   \right] m_{\mathrm{I}}^2
%                  \right. \nonumber\\
%&&\left. +  %   \frac{g}{4\pi^2}
%        y \int\left[
% ({y^2-3})/{\sqrt{y^2-1}}
%+3\mbox{ch}^{-1}(y)/y
%                     \right]
%       dm_{\mathrm{I}}^2
%                     \right\}
%                 \nonumber\\
%&& +\ \mbox{higher order terms}\ %  m_{\mathrm{I}}
%\end{eqnarray}
%
%with
%$y\equiv \mu/m_0$.
%
where
$
P_{\mathrm{id}}
\equiv
{g}/{(48\pi^2)}
 [\mu\sqrt{\mu^2-m_0^2}\ (2\mu^2-5m_0^2)
    +3m_0^4\mbox{ch}^{-1}(\mu/m_0)
 ]
 $
is the pressure of a degenerate non-interacting quark
plasma. The convergence of Eq.\ (\ref{Pexpand})
can be mathematically proven. Now substituting
%Eq.~(\ref{mIpert})
$m_{\mathrm{I}}=\alpha_0\mu
=\sqrt{2\alpha_{\mathrm{s}}/\pi}\mu$
into the above perturbative expansion,
and then taking the limit of $m_0\rightarrow 0$ due to
the extreme smallness of the current mass of light quarks,
the second term vanishes while the first and third terms
give
$ %\begin{equation}
{P}/{P_{\mathrm{id}}}
=1-2{\alpha_{\mathrm{s}}}/{\pi}
$ %\end{equation}
which is consistent with the hard-thermal-loop
resumed pressure \cite{Baier00PRL} and
the weak-coupling expansion
\cite{Freedman78PRD,Toimela85IJTP}.

At the same time, the second term of Eq.\ (\ref{mIexpn})
dominates at higher densities,
 i.e.,
$ %\begin{equation} \label{mIhigh}
m_{\mathrm{I}}
=a_1p_{\mathrm{f}}.
$ %\end{equation}
So inserting
$p_{\mathrm{f}}=m_{\mathrm{I}}/a_1$ into Eq.\ (\ref{pfmu})
and solving for $m_{\mathrm{I}}$ give
$
m_{\mathrm{I}}
=a_1[\sqrt{(1+a_1^2)\mu^2-m_0^2}-a_1m_0]/(1+a_1^2)
$
which approaches, at the limit of $m_0\rightarrow 0$,
 to $m_{\mathrm{I}}=a_1\mu/\sqrt{1+a_1^2}$.
Comparing this with
%  Eq.\ (\ref{mIpert})
$m_{\mathrm{I}}=\alpha_0\mu$
 gives $a_1/\sqrt{1+a_1^2}=\alpha_0$. Solving for $a_1$
from this equality then leads to Eq.\ (\ref{C1alphas}).

In order to perform the integration in Eq.\ (\ref{nbexp}),
one should know the total derivative of Eq.\ (\ref{mIexpn}), i.e.,
\begin{equation} \label{dmIdp}
\frac{dm_{\mathrm{I}}}{dp_{\mathrm{f}}}
=\frac{\partial m_{\mathrm{I}}}{\partial p_{\mathrm{f}}}
 +\frac{\partial m_{\mathrm{I}}}{\partial a_1}
  \frac{da_1}{dp_{\mathrm{f}}}
=-\frac{a_{-1}}{p_{\mathrm{f}}^2}+a_1
 +p_{\mathrm{f}}\frac{da_1}{dp_{\mathrm{f}}}.
\end{equation}
From Eqs.\ (\ref{pfmu}), (\ref{C1alphas}),
and the Gell-Mann-Low equation
$ %\begin{equation}
\mu{d\alpha_{\mathrm{s}}}/{d\mu}
=\sum_{n=0}^{\infty}C_n\alpha_{\mathrm{s}}^{n+2}
       \label{GLeq}
$ %\end{equation}
\cite{GellmannLow}, it can be shown that the derivative
of $a_1$ with respect to  $p_{\mathrm{f}}$ is
\begin{equation} \label{daldp}
\frac{da_1}{dp_{\mathrm{f}}}
=\frac{p_{\mathrm{f}}+m\partial m/\partial p_{\mathrm{f}}
      }
        {
 \frac{d\alpha_{\mathrm{s}}/da_1}
      {\sum_{n=0}^{\infty}C_n\alpha_{\mathrm{s}}^{n+2}}
 (p_{\mathrm{f}}^2+m^2)
 -m\frac{\partial m}{\partial a_1}
        }
\end{equation}
with
$
\partial m/\partial p_{\mathrm{f}}
=-a_{-1}/p_{\mathrm{f}}^2+a_1,
$
$
\partial m/\partial a_1
=p_{\mathrm{f}},
$
$
d\alpha_{\mathrm{s}}/da_1
=\pi a_1/(1+a_1^2)^2.
$
To second order, the Gell-Mann-Low equation can be
integrated out, giving
\begin{equation}  \label{asmu}
 \frac{1}{\alpha_{\mathrm{s}}}
-\frac{1}{\alpha_{\mathrm{s}}(1)}
+\frac{C_1}{C_0}\ln
 \frac{C_1+C_0/\alpha_{\mathrm{s}}(1)}
      {C_1+C_0/\alpha_{\mathrm{s}}}
+C_0\ln\frac{\mu}{\Lambda}
=0,
\end{equation}
where $\alpha_{\mathrm{s}}(1)$ is the value
of $\alpha_{\mathrm{s}}$ at the QCD scale point $\mu=\Lambda$,
$
C_0=-(11N_{\mathrm{c}}-2N_{\mathrm{f}})/(6\pi)
$
and
$
C_1
=-
 [(34N_{\mathrm{c}}-13N_{\mathrm{f}})N_{\mathrm{c}}
  +3N_{\mathrm{f}}/N_{\mathrm{c}}]/(24\pi^2)
$
are the Gell-Mann-Low coefficients \cite{GLxisu}.
The scale parameter $\Lambda$\ is usually taken to be 300 MeV
while $\alpha_{\mathrm{s}}(1)$ is taken to be 1 \cite{Freedman78PRD}.
The confinement parameter $a_{-1}$ should satisfy the constraint:
the energy per baryon for the two-flavor case is no less than
930 MeV, in order not to contradict standard nuclear physics
\cite{Peng00PRC61}. In the present calculation, $a_{-1}$ is taken
to be such a value that the maximum value of the QCD running
coupling $\alpha_{\mathrm{s}}$ just does not exceed 1 at whole
densities. If larger $a_{-1}$ values are used, the condensate
goes to zero more rapidly.

  For a given $p_{\mathrm{f}}$, the running coupling
$\alpha_{\mathrm{s}}$ is obtained by solving Eq.~(\ref{asmu})
with the help of Eqs.\ (\ref{pfmu}), (\ref{mIexpn}),
and (\ref{C1alphas}).
Inserting Eq.\ (\ref{dmIdp}) with Eq.\ (\ref{daldp}) into
Eq.\ (\ref{nbexp}) will give the baryon number density
$n_{\mathrm{b}}$.
The chiral condensate can then be calculated from
Eq.\ (\ref{qc2}).

%At higher densities and the limit of $m_0\rightarrow 0$,
%Eq.\ (\ref{dmIdp}) becomes
%$
%dm_{\mathrm{I}}/dp_{\mathrm{f}}
%=a_1-(1+a_1^2)
% /[a_1-5\pi a_1/((1+a_1^2)^2
%  \sum_{n=0}^{\infty}C_n \alpha_{\mathrm{s}}^{n+2})]
%\equiv \alpha_{\prime}.
%$
%Substituting this and
% $m_{\mathrm{I}}=a_1p_{\mathrm{f}}$ % and (\ref{daldp})
%into Eq.~(\ref{nbexp}) gives
%$ %\begin{equation}
%p_{\mathrm{f}}
%=\left[
% ({18\pi^2}{g}) {n_{\mathrm{b}}}/{\alpha^*}
% \right]^{1/3}
%$ %\end{equation}
%%
%%\begin{equation}
%%\mbox{with}\
%%\alpha^*
%%=1+
%% \frac{\alpha_0^2}{2(1-\alpha_0^2)}
%% \left[
%% 1-\frac{\alpha_0^2\mbox{arcsech}\left(\alpha_0\right)}
%%        {\sqrt{1-\alpha_0^2}}
%% \right].
%%\end{equation}
%with
%$%\begin{equation}
%\alpha^*=1
%+\frac{1}{2}\alpha^{\prime} a_1
% [1+a_1^2\mbox{arcsinh}(1/a_1)/\sqrt{1+a_1^2}].
%$\ %\end{equation}
%Correspondingly, Eq.~(\ref{qc2}) reads
%\begin{eqnarray}
% &
%{\langle\bar{q}q\rangle_{n_{\mathrm{b}}}}
%     /{\langle\bar{q}q\rangle_0}
%= 1-{n_{\mathrm{b}}}/{n_{\mathrm{h}}}
% &                    \label{qcpert} \\
%\mbox{with} \ \
%&
%n_{\mathrm{h}}
%\equiv
%n^*/[F(1/a_1)/\alpha^*-(3/4){\alpha^*}^{1/3}/a_1].
%&
%\end{eqnarray}
%Here the approximations $m\approx m_{\mathrm{I}}$
%and $F(p_{\mathrm{f}0}/m_0)\approx (3/4)p_{\mathrm{f}0}/m_0$
%can be used.

Now let's find the lower density behavior of Eq.~(\ref{qc2}).
% which serves also as an account for the validity of
% the first term in Eq.\ (\ref{mIexpn}).
Because $m\approx m_{\mathrm{I}}\gg p_{\mathrm{f}}$
and $F(p_{\mathrm{f}}/m)\approx 1$ at lower densities,
Eqs.\ (\ref{dEIexp}) and (\ref{EIexp}) become
$m_{\mathrm{I}}=dE_{\mathrm{I}}/(3dn_{\mathrm{b}})$
and
$
(p_{\mathrm{f}}/p_{\mathrm{f}0})^{3}m_{\mathrm{I}}
=E_{\mathrm{I}}/(3n_{\mathrm{b}}).
$
Denoting the inter-quark interaction by $\mbox{v}(\bar{r})$
with
 $
\bar{r}
%=(2\pi^{-1}/n_{\mathrm{b}})^{1/3}
\propto 1/n_{\mathrm{b}}^{1/3}
$
being the average distance of quarks \cite{Peng02PLB548}.
The interacting energy per baryon
$E_{\mathrm{I}}/(3n_{\mathrm{b}})$ is proportional to
$\mbox{v}(\bar{r})$. Because the confinement interaction dominates
at lower densities while the confinement is linear according to
lattice simulations \cite{lattice}, one has
$ % \begin{equation}
{E_{\mathrm{I}}}/{(3n_{\mathrm{b}})}
\propto \bar{r}^Z.\
$%\end{equation}
Here the confinement exponent is denoted by $Z$.
For linear confinement, it is equal to unity.
With a proportion coefficient $\sigma$, one can write
$ %\begin{equation} \label{Eisimp}
E_{\mathrm{I}}
=3\sigma n_{\mathrm{b}}^{1-Z/3}.
$ %\end{equation}
%
%It will soon be seen that the concrete value of
%the coefficient $\sigma$ does not influence
%the in-medium condensate at lower densities.
%
Substituting this into
the approximate equalities of
Eqs.~(\ref{dEIexp}) and (\ref{EIexp})
just obtained
 gives
$ %\begin{eqnarray}
m_{\mathrm{I}}
={\sigma(1-Z/3)}/
        {n_{\mathrm{b}}^{Z/3}}  \label{mIsimp}
$
and
$
p_{\mathrm{f}}
=    [
   ({18\pi^2}/{g})
   n_{\mathrm{b}}/(1-Z/3)
     ]^{1/3}.       \label{pfsimp}
$ %\end{eqnarray}
These two equalities give an account for the
validity of the first term in Eq.\ (\ref{mIexpn}).
They also show that the effective Fermi momentum
has been boosted to a higher value by a factor,
i.e., $p_{\mathrm{f}}^3/p_{\mathrm{f}0}^3=1/(1-Z/3)$.
This ratio can also be obtained by expanding the integrand of the
integration in Eq.\ (\ref{nbexp}) to Taylor series at
$p_{\mathrm{f}}=0$, taking the leading term
$2p_{\mathrm{f}}^3/(3m)$, making the variable substitution
$m_{\mathrm{I}}=\alpha_{-1}/p_{\mathrm{f}}^Z$
[Z=1 corresponds to the first term of Eq.\ (\ref{mIexpn})],
and then completing the integration.
Finally substituting the ratio
% and the approximations
%$m_{\mathrm{I}}\approx m\gg p_{\mathrm{f}}$
%and $F(p_{\mathrm{f}}/m)\approx 1$
%% Eqs.~(\ref{mIsimp}) and (\ref{pfsimp})
into Eq.~(\ref{qc2}) gives
$ %\begin{equation} \label{qcL}
{\langle\bar{q}q\rangle_{n_{\mathrm{b}}}}
     /{\langle\bar{q}q\rangle_0}
= 1-{n_{\mathrm{b}}}/{n_{\mathrm{l}}}
\ \mbox{with} \
n_{\mathrm{l}}
=(1-Z/3)n^*.
$ %\end{equation}

Now demanding a compatibility with the famous
model-independent result in nuclear matter, i.e.,
$ %\begin{equation} \label{qcnlin}
{\langle\bar{q}q\rangle_{\rho}}/{\langle\bar{q}q\rangle_0}
=1-{\rho}/{\rho^*}
 \ \mbox{with} \
\rho^*\equiv {M_{\pi}^2F_{\pi}^2}/{\sigma_{\mathrm{N}}}
$ %\end{equation}
\cite{Cohen92PRC45},
%which was first proposed by Drukarev {\sl et al.}\
%\cite{Drukarev99}, and later re-justified by many authors
%\cite{Cohen92PRC45,Chanfray93,Lutz},
one can get, from $n_{\mathrm{l}}=\rho^*$,
an interesting relation
\begin{equation} \label{sigm0}
\sigma_{\mathrm{N}}
=\frac{9m_0}{3-Z}
\end{equation}
which relates the pion-nucleon sigma term
$\sigma_{\mathrm{N}}$, the confinement exponent $Z$,
and the average current mass of $u/d$ quarks whose value
is about $m_0=7.5$ MeV.
In obtaining the above equation,
the Gell-Mann-Oakes-Renner relation
$-2m_0\langle\bar{q}q\rangle_0=M_{\pi}^2F_{\pi}^2$
has been applied.
% $m_0=(m_{u0}+m_{d0})/2=(5+10)/2=7.5$ MeV. % \cite{Gasser82}.
For the linear confinement of $Z=1$,  Eq.\ (\ref{sigm0}) gives
$\sigma_{\mathrm{N}}=\frac{9}{2}m_0\approx 34$ MeV.
The reported $\sigma_{\mathrm{N}}$ values in literature are
significantly different.
Previously, it was reported to be a smaller one, e.g., in the range
of 25---26 MeV \cite{Banerjee77}. Later, a bigger value, e.g.,
$\sigma_{\mathrm{N}}\approx 56.9\pm 6.0$ MeV \cite{Gensini80} was arrived at.
After that, modest values of about 45 MeV was obtained
\cite{Ericson87,Gasser91}. The accepted value used to be as high as
around 65 MeV \cite{Sainio95}. Recently, a different value of about 50 MeV
was reported again \cite{Gibbs98}.
This phenomena is due to the fact that $\sigma_{\mathrm{N}}$ is not
directly measurable but extrapolated according to special models.
Obviously, the $\sigma_{\mathrm{N}}$ value here is
bigger than previously determined, but smaller than
presently accepted. Naturally, if one would like to reproduce
the bigger values, the confinement would have
to be also bigger. However, lattice simulations favor
the linear confinement \cite{lattice}.

 \begin{figure}[htb]
\begin{center}
 \scalebox{.65}{\includegraphics{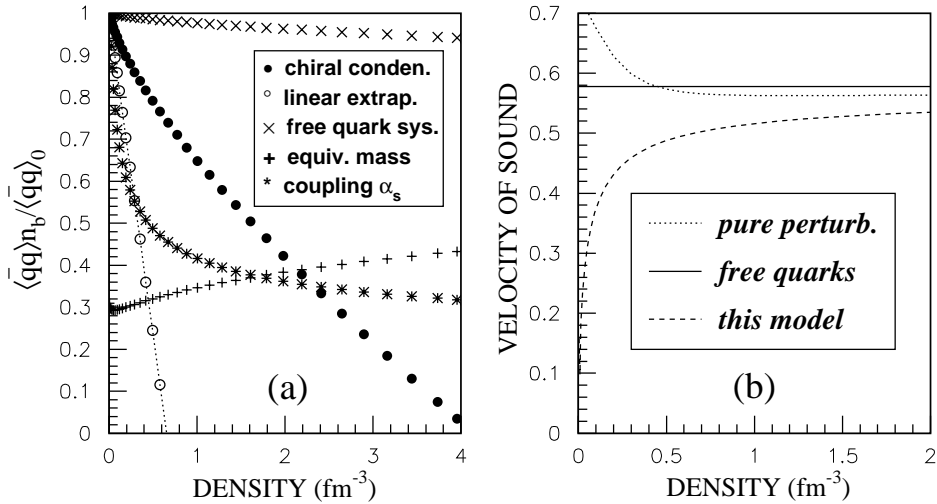}}
% \epsfxsize=12cm
% \epsfbox{fig1.eps}
\end{center}
 \caption{
(a) Density dependence of the in-medium chiral condensate.
It is generally a decreasing function of the density (star line),
approaching to zero at about 4 fm$^{-3}$.
The open-circle line on the left is the linear extrapolation.
The up-most line with crosses is for a quark system without interactions.
The plus-marked line gives the corresponding equivalent mass,
in unit of the nucleon mass 939 MeV.
The star line is the QCD coupling $\alpha_{\mathrm{s}}$.
(b) Velocity of sound in the pure perturbative calculation
(the dotted line) and in the present approach (the dashed line).
         }
 \label{fig1}
 \end{figure}

%Numerical results are plotted in Fig.~\ref{fig1}.
%The up-most line is for the chiral condensate in a free quark
%gas. % drawn from Eq.~(\ref{qcfree}).
%Because the decreasing
%speed is proportional to the current mass which is very small,
%the line decreases slowly. When interactions are considered,
%the condensate goes down much more rapidly, approaching to
%zero at about 4 fm$^{-3}$ which is still an extremely high density.
%The linear behavior % of Eq.~(\ref{qcL})
%is shown with a triangle line on the left.
%The equivalent mass is given, in unit of the nucleon
%mass $M_{\mathrm{N}}=939$ MeV, with a plus-marked line,
%which shows that the ratio $m/M_{\mathrm{N}}$ is about $0.2$.
%This result can also be understood within hadronic degree of freedom.
%At lower densities, the energy density of nuclear matter
%is near $M_{\mathrm{N}}\rho$\ with $\rho$\ being the nucleon density
%which is one third of the total quark number density.
%Replacing the left hand side of Eq.\ (\ref{epsilon})
%with $M_{\mathrm{N}}\rho$\ and considering
%$F(p_{\mathrm{f}}/m) \sim 1$
%and
%$p_{\mathrm{f}}^3 \sim p_{\mathrm{f0}}^3/(1-Z/3)$,
%one has $3m\approx (1-Z/3)M_{\mathrm{N}}$. Taking $Z=1$ then gives
%$m\approx (2/9)M_{\mathrm{N}}$.
%In constituent quark models, the constituent quark
%mass is usually taken to be $(1/3)M_{\mathrm{N}}$.
%From the present result, it seems more reasonable to
%take a little smaller value of $(2/9)M_{\mathrm{N}}$,
%due to medium effect.

Numerical results are plotted in Fig.~\ref{fig1}.
In Fig.~\ref{fig1} (a),
%the in-medium chiral condensate (full-circle line)
%decreases with increasing densities,
%approaching to zero at about 4 fm$^{-3}$.
the up-most line is for the chiral
condensate in a free quark gas. % drawn from Eq.~(\ref{qcfree}).
Because the decreasing
speed is proportional to the current mass which is very small,
the line goes down slowly. When interactions are considered,
the condensate (full circles) decreases much more rapidly, approaching to
zero at about 4 fm$^{-3}$ which is still an extremely high density.
The linear behavior % of Eq.~(\ref{qcL})
is shown with an open-circle line on the left.
The equivalent mass is given, in unit of the nucleon
mass $M_{\mathrm{N}}=939$ MeV, with a plus-marked line,
which shows that the ratio $m/M_{\mathrm{N}}$ is about 0.3--0.4.
In constituent quark models \cite{zyzhang1997NPA625}, the constituent
quark mass is usually taken to be $M_{\mathrm{N}}/3$,
consistent with the present result.

Equations (\ref{Pmu}) and (\ref{epsilon}) provide the equation of
state. If the energy per baryon $E/n_{\mathrm{b}}$ is plotted as
a function of the density $n_{\mathrm{b}}$, the minimum point
should correspond exactly to the zero pressure,
which is a general requirement of thermodynamics,
as has be shown in Ref.\ \cite{Peng00PRC62}. However, this is not
the case in the pure perturbative results, e.g., a check of the
figure 7 in Ref.\ \cite{Freedman78PRD} shows that the density
corresponding to zero pressure is smaller than the density
corresponding to the minimum energy per quark.
This has a strong consequence on the velocity of sound,
calculated by $|dP/dE|^{1/2}$, which has been
shown in Fig.~\ref{fig1} (b) for the present approach (the dash line)
and in the weak expansion (the dotted line).
At higher densities, they approach asymptotically to the
ultra-relativistic case (the full line) as expected.
However, their lower density behavior is completely opposite.
This may have some stringent consequences on dynamic situations,
such as the possibility of strange candidates in neutron stars
or perhaps heavy ion collisions.

  In summary, the density behavior of the in-medium chiral
condensate has been studied with a new approach.
 The chiral condensate is shown to be generally a decreasing function
of density. At lower densities, it decreases linearly,
which means that the pion-nucleon sigma term is 9/2 times
the average current mass of light quarks if the confinement
is linear. With increasing densities, the deviation from
linear extrapolation becomes significant.
Considering both perturbative and non-perturbativre effects,
the condensate approaches to zero at about 4 fm$^{-3}$.

%\vspace{0.5cm}
\section*{Acknowledgements}

The author would like to thank support from
the DOE (DF-FC02-94ER40818),
NSFC (10375074, 90203004, 19905011),
FONDECYT (3010059 and 1010976),
CAS (E-26), and BES-BEPC.
He also acknowledges hospitality at MIT-CTP.

\end{document}